\documentstyle[12pt]{article}
\begin{document}
\baselineskip=26pt
  
 
 
\vspace{0.3in}
 
\centerline{Oscillations of recoil particles against mixed states}
   
\vspace{0.2in}
 
\centerline{H. Burkhardt\footnote{Also 
at Shell Centre for Mathematical Education, University, Nottingham NG7 2RD,
England}, J. Lowe\footnote{Also at Physics Department, University, 
Birmingham B15 2TT, England}, G.J. Stephenson Jr.}
 
\centerline{\it Physics Department, University of New Mexico, Albuquerque, 
NM 87131, USA}
  
\vspace{0.1in}
  
\centerline{and}
 
\vspace{0.1in}
 
\centerline{T. Goldman}
 
\centerline{\it Theoretical Division, Los Alamos National Laboratory, 
Los Alamos, NM 87545, USA}
 
\vspace{0.5in}
  
ABSTRACT: Some consequences of the oscillations of neutral 
kaons and neutrinos are discussed, in particular, the possibility of 
oscillations of particles recoiling against kaons or neutrinos from 
the production process. We show that there are no stationary oscillations 
of these recoil particles in any order, and that the apparent 
long-wavelength oscillations, which might appear to result when an earlier 
treatment of ours was taken to higher order, are spurious. We show that 
the recoil particles may show a travelling interference pattern. It may
be possible to observe this pattern for $\Lambda$s produced in a 
reaction, but there seems to be little hope of observing this for the 
case of neutrinos from muon decay.
 
\vspace{1.0in}
 
\centerline{\bf I. Introduction}
 
\vspace{0.2in}
 
The subject of the oscillations of particles produced in 
a mixture of mass eigenstates has been discussed for many years. 
An early example is that of the neutral kaons. Hadronic reactions
produce these in strangeness eigenstates, $K^0$ and $\bar{K}^0$, 
which are mixtures of the mass eigenstates, $K_L$ and $K_S$. 
The possibility has also been discussed that neutrinos might
show similar characteristics. It may be that the familiar flavour
eigenstates, $\nu_e$, $\nu_{\mu}$ and $\nu_{\tau}$  are mixtures
of mass eigenstates, $\nu_1$, $\nu_2$ and $\nu_3$. If so, they
should show a similar behaviour to that of neutral kaons.
 
\vspace{0.1in}
 
The quantum mechanics of the oscillations of such a system have been
treated in many standard texts[1]. If a kaon is produced initially in
a pure $K^0$ or $\bar{K}^0$ state, the system  oscillates between a 
$K^0$ and a $\bar{K}^0$, approaching the equal mixture of a pure 
$K_L$ state. This is the well-known phenomenon of strangeness 
oscillations[1]. Similar oscillations may occur for neutrinos
produced in one of the flavour eigenstates, $\nu_e$, $\nu_{\mu}$ or
$\nu_{\tau}$. 
 
\vspace{0.1in}
 
For some years, the oscillations of a neutral kaon or a neutrino
were treated in isolation, without regard for the details of the 
reaction in which the particles are produced. Recently, a series of 
papers was published by Srivastava, Widom and Sassaroli[2,3,4] in which 
the kinematics of the production process are considered in detail. 
They point out that, in the usual experimental situations, the mass 
eigenstates ($K_L$ and $K_S$ for kaons, $\nu_1$, $\nu_2$ and $\nu_3$ 
for neutrinos) have different momenta {\it and} different energies, 
and they investigated the consequences of this for both the neutral 
particle and the other particles in the reaction.
 
\vspace{0.1in}
 
In examining the calculation of Srivastava {\it et al.}, we found[5]
some errors in the treatment of the relation between coordinates in
the various frames. On correcting these, we found results for the
oscillation pattern that are different from those of ref. [2], and
are, in fact, consistent with the conventional description of kaon
or neutrino oscillations. This is surprising, because we incorporated
the detailed treatment of the kinematics of the production process,
pointed out by Srivastava {\it et al.} and by Goldman[6], and earlier by 
Boehm and Vogel[7]. 
It seems that this does not change the observed oscillation pattern. 
Our results have since been confirmed by other authors[8,9].
 
\vspace{0.1in}
 
One of the important differences between the results of ref. [2] and
those of refs. [5,8,9] is that, in the former, oscillations in the
strangeness of the neutral kaon are predicted to give rise to 
oscillations in the $\Lambda$s produced in association with the
kaon. No such effect was predicted in refs. [5,8,9] or in any 
earlier work. The possible existence of such oscillations in the 
neutrino case would be of immense practical importance; the 
experimentally challenging problem of detection of neutrino 
oscillations could be replaced by the much simpler problem of 
observation of, for example, muons from the decay $\pi\rightarrow\mu\nu$. 
Partly because of this, we have extended our earlier calculation, 
and in the present paper, we examine two aspects of the problem:
 
\noindent (a) In ref. [5], in common with all treatments so far
published, the calculations are carried out only to first order in
the mass difference $\delta m = m_L-m_S$. In fact, if our calculation
is continued to higher orders, oscillations of the recoil particle 
appear. However, these are spurious; in section II we prove that in a 
full calculation there are no oscillations, and we show how these 
spurious oscillations appear. 
 
\noindent (b) Because of the substantial importance of neutrino
oscillation experiments, we searched for a possible experimental
configuration in which neutrino oscillations could be inferred from
measurements on the recoiling muon. There is no stationary 
oscillation pattern in the lab frame. However, a travelling
oscillation pattern may exist. In section III we discuss some aspects
of the wave packet solutions for these systems, and in section IV, we 
discuss the possibility of observing the travelling oscillation pattern. 
We conclude that, although it may be possible to observe this pattern 
for $\Lambda$s produced in a reaction, it is not at present possible 
in the case of neutrino oscillations.
   
\vspace{0.2in}
 
\centerline{\bf II. Exact treatment of kaon oscillations}
 
\vspace{0.2in}
 
We start by summarising the results of our earlier paper [5]. As in
that paper, we consider the case of a neutral kaon produced in the
reaction 
 
$$\pi^- p\rightarrow \Lambda K^0.$$
 
\noindent The equations for neutrinos produced in a flavour eigenstate 
by a weak process require the treatment of at least three mass 
eigenstates. We do not consider the additional complications here, 
but we treat the case of two coupled mass eigenstates for which the 
equations are analogous. The kaon state at the moment of production, 
therefore, is the strangeness 1 meson $K^0$, which is a mixture of the 
mass eigenstates
 
\begin{eqnarray}
\mid K^0\rangle=\sqrt{\frac{1+\mid\epsilon\mid^2}{2(1+\epsilon)^2}}
\left(\mid K_S\rangle+\mid K_L\rangle\right)
\end{eqnarray}
 
\noindent where $\epsilon$ is the usual CP-violation parameter. Thus
the $\Lambda K$ state at the moment of production, $t=0$, is
 
\begin{eqnarray}
\mid\Lambda K(t=0)\rangle=\sqrt\frac{1+\mid\epsilon\mid^2}{2(1+\epsilon)^2}
~~\{\mid\Lambda_SK_S\rangle+\mid\Lambda_LK_L\rangle\}.
\end{eqnarray}
 
\noindent The essential point here is that $K_L$ and $K_S$ are two 
different particles with with different masses {\it and} different 
momenta. Thus the recoiling $\Lambda$s produced in association with 
them have different momenta. We denote these $\Lambda$ states by 
$\Lambda_L$ and $\Lambda_S$ respectively; they are, of course, the 
same particle. Denoting the total center-of-mass energy by $\sqrt{s}$, 
the center-of-mass momenta of the $K_L$, $K_S$, $\Lambda_L$ and 
$\Lambda_S$ are given by

$$p_i^2=\frac{(s-m_i^2-m_{\Lambda}^2)^2-4m_i^2m_{\Lambda}^2}{4s}$$

\noindent where $i=L$ or $S$. 
 
\vspace{0.1in}
 
The state (2) develops in time according to
 
\begin{eqnarray}
\mid\Lambda K(t)\rangle=\sqrt\frac{1+\mid\epsilon\mid^2}{2(1+\epsilon)^2}
~~\{a_S(\tau_{\Lambda_S},\tau_{K_S})\mid\Lambda_SK_S\rangle+
a_L(\tau_{\Lambda_L},\tau_{K_L})\mid\Lambda_LK_L\rangle\}
\end{eqnarray}
 
\noindent where
 
\begin{eqnarray}
a_i(\tau_{\Lambda_i},\tau_{K_i})={\rm exp}
\{-i(m_{K_i}\tau_{K_i}+m_{\Lambda}\tau_{\Lambda_i})-
\frac{1}{2}(\Gamma_{K_i}\tau_{K_i}+\Gamma_{\Lambda}\tau_{\Lambda_i})\}
\end{eqnarray}
 
\noindent with $i=S$ or $L$. The four proper times, $\tau_{K_L}$, 
$\tau_{K_S}$, $\tau_{\Lambda_L}$ and $\tau_{\Lambda_S}$, are related to 
the time in the overall center-of-mass frame, $t$, by the appropriate 
Lorentz transformations relating a point $(\xi_i,\tau_i)$ in the frame 
$i$ to the point $(x,t)$ in the overall center-of-mass frame:
 
\begin{eqnarray}
\xi_i=\gamma_i(x-\beta_it)
\end{eqnarray}
 
\begin{eqnarray}
\tau_i=\gamma_i(t-\beta_ix).
\end{eqnarray}
 
\noindent A crucial point in our paper[5] is that the two times 
$\tau_{K_L}$ and $\tau_{K_S}$ are related to a {\it single} space-time 
point, $(x_K,t_K)$, in the overall c.m. system, and similarly for 
the times  $\tau_{\Lambda_L}$ and $\tau_{\Lambda_S}$ for the $\Lambda$.
 
\vspace{0.1in}
 
Although the analysis is carried out in plane-wave formalism, a more
realistic solution in a practical case involves wave packets constructed
from these plane waves. Construction of wave packets is discussed in
more detail below; for now, we need only assume that the wave packets
will be large enough that the $K_L$ and $K_S$ packets do not separate
significantly, but are nevertheless small compared with the dimensions 
of the apparatus. Similar statements hold for the $\Lambda$ momentum 
components, $\Lambda_L$ and $\Lambda_S$. Further, the packet will be 
centered approximately on the position of the classical particle. 
This condition specifies a relation between the time and position of 
observation of a particle. If we choose to observe at a space point $x$, 
then the time of observation must be $t=x/\beta$, where $\beta$ is the 
classical particle velocity. The velocity $\beta$ could be chosen to be 
$\beta_L$ or $\beta_S$ or some average of these. In our previous work[5], 
we used $\overline{\beta}$, the average of $\beta_L$ and $\beta_S$. 
Here, we leave the choice of $\beta$ until later; it will turn out to 
be important, but for now, we merely require that it does not differ 
appreciably from $\beta_L$ or $\beta_S$.
 
\vspace{0.1in}
 
The derivation now proceeds as in ref. [5]. Since $t=x/\beta$, the 
Lorentz transformations (6) give
 
\begin{eqnarray}
\tau_i=\gamma_i\left(\frac{x}{\beta}-\beta_ix\right)=
\gamma_ix\left(\frac{1}{\beta}-\beta_i\right).
\end{eqnarray}
 
\noindent Then (4) becomes
 
\begin{eqnarray}
a_i(t)={\rm exp} \left[-i\left(m_{\Lambda}\gamma_{\Lambda_i}
(\frac{1}{\beta_{\Lambda}}-
\beta_{\Lambda_i})x_{\Lambda}+m_{K_i}\gamma_{K_i}(\frac{1}
{\beta_K}-\beta_{K_i})x_K\right)\right.
\nonumber \\
-\left.\frac{1}{2}\left(\Gamma_{\Lambda}\gamma_{\Lambda_i}(\frac{1}
{\beta_{\Lambda}}-
\beta_{\Lambda_i})x_{\Lambda}+\Gamma_{K_i}\gamma_{K_i}
(\frac{1}{\beta_K}-\beta_{K_i})x_K\right)\right]
\end{eqnarray}
 
\noindent and the state vector at center-of-mass time $t$ is
 
\begin{eqnarray}
\mid\Lambda K(t)\rangle=\sqrt\frac{1+\mid\epsilon\mid^2}{2(1+\epsilon)^2}
~~\{a_S(t)\mid\Lambda_SK_S\rangle+a_L(t)\mid\Lambda_LK_L\rangle\}.
\end{eqnarray}
 
\vspace{0.1in}
 
\noindent As in ref.[5], we project out a specific strangeness for 
the kaon. If we choose strangeness 1, we get
 
\begin{eqnarray}
\psi_{\Lambda K^0}(x_{K^0},x_{\Lambda})&=&\sqrt\frac{1+\mid\epsilon\mid^2}
{2(1+\epsilon)^2}~~\{a_S(t)\langle\Lambda K^0\mid\Lambda_SK_S\rangle+a_L(t)
\langle\Lambda K^0\mid\Lambda_LK_L\rangle\}
\nonumber \\
&=&\frac{1}{2}\{a_S(t)+a_L(t)\}
\end{eqnarray}
 
\vspace{0.1in}
 
\noindent and the $\bar{K}^0$ part has the opposite sign in the bracket,
$\{a_S(t)-a_L(t)\}$. Writing $a_i(t)$ as $a_i(t)={\rm exp}(-ib_i-c_i)$,
the joint probability distribution for detection of a $K^0$ and a 
$\Lambda$ is given by
 
\begin{eqnarray}
P(x_{\Lambda},x_{K^0})&=& \frac{1}{4}\mid a_S(t)+a_L(t)\mid^2
\nonumber \\
&=&\frac{1}{4}\{\mid a_S(t)\mid^2+\mid a_L(t)\mid^2+
2{\rm e}^{-(c_S+c_L)}{\rm cos}(b_L-b_S)\}
\nonumber \\
&=&\frac{1}{4}\{{\rm e}^{-2c_L}+{\rm e}^{-2c_S}+
2{\rm e}^{-(c_S+c_L)}{\rm cos}(b_L-b_S)\}.
\end{eqnarray}
 
\noindent The oscillations arise from the cosine term, 
${\rm cos}(b_L-b_S)$, where
 
\begin{eqnarray}
b_i=m_{\Lambda}\gamma_{\Lambda_i}(\frac{1}{\beta_{\Lambda}}-
\beta_{\Lambda_i})x_{\Lambda}+m_{K_i}\gamma_{K_i}(\frac{1}{\beta}-
\beta_{K_i})x_K.
\end{eqnarray}
 
\vspace{0.1in}
 
In ref. [5], the quantity $(b_L-b_S)$ was evaluated to first order in 
$\delta m$ from the relation $b_L-b_S=(\partial b/\partial m)\delta m$. 
The result was that $(b_L-b_S)$ is a function of $x_K$ but not of
$x_\Lambda$, showing that in spite of a more detailed treatment of
the kinematics than in conventional derivations, there are no 
oscillations of the $\Lambda$ to first order. We also found that
the kaon oscillations have the same wavelength as in the usual derivation,
not as in ref.[2]. The possibility that oscillations exist in higher 
order is of considerable experimental importance for the case of 
neutrino oscillations. If their wavelength were not prohibitively long, 
then detection of oscillations of a recoil particle, for example muons 
from the $\pi\rightarrow\mu\nu$ decay, might provide an indirect, 
relatively simple method for studying neutrino oscillations. However, 
we shall show that such oscillations cannot exist in any order. 
 
\vspace{0.1in}
 
To prove this, observe that the coefficient of $x_{\Lambda}$ in the 
expression for $b_L-b_S$ is
  
$$m_{\Lambda}\gamma_{\Lambda_L}\left(\frac{1}{\beta}-
\beta_{\Lambda_L}\right)-
m_{\Lambda}\gamma_{\Lambda_S}\left(\frac{1}{\beta}-
\beta_{\Lambda_S}\right).$$
 
\noindent Setting this equal to zero, we can solve for $\beta$. We
denote the resulting value by $\beta^*$, and we find
 
\begin{eqnarray}
\beta^*_{\Lambda}=\frac{E_{\Lambda_L}-E_{\Lambda_S}}
{p_{\Lambda_L}-p_{\Lambda_S}}=\frac{p_{\Lambda_L}+p_{\Lambda_S}}
{E_{\Lambda_L}+E_{\Lambda_S}}.
\end{eqnarray}
 
\noindent We denote by $S^*$ the frame defined by the velocity 
$\beta^*$. The $S^*$ frame is the c.m. frame of the two components of 
the $\Lambda$, i.e. the frame in which $\Lambda_L$ and $\Lambda_S$ 
have equal energy and opposite momenta. Thus, to the extent that one
may choose $\beta^*$ rather than $\bar{\beta}$ to define the time
of observation, one may prove that the coefficient of $x_{\Lambda}$ 
vanishes exactly, and hence there can be no stationary oscillations of 
the $\Lambda$s from $\pi^- p\rightarrow\Lambda K^0$ nor, by analogy, of 
muons from $\pi\rightarrow\mu\nu$.
  
\vspace{0.1in}
 
For the kaon, we can again define a frame in which the two components 
have equal and opposite momenta, by
 
\begin{eqnarray}
\beta^*_K=\frac{p_{K_L}+p_{K_S}}{E_{K_L}+E_{K_S}}.
\end{eqnarray}
 
\noindent In this frame, the $S^*$ frame for the kaon, the coefficient 
of $x_K$ in the expression for $(b_L-b_S)$ is exactly
 
$$\frac{m_L^2-m_S^2}{p_L+p_S}.$$
 
\noindent To first order in $\delta m$, this is $m\delta m/p$, in 
agreement with our previous work[5] and with the standard result[1].
 
\vspace{0.1in}
 
The origin of the apparent long-wavelength $\Lambda$ 
oscillations that appear if the treatment of ref.[5] is taken to 
higher orders is now clear. In their corresponding $S^*$ frames, 
$p^*_{\Lambda_S}=-p^*_{\Lambda_L}$ and $p^*_{K_S}=-p^*_{K_L}$. The 
energies are equal for the $\Lambda$, i.e. 
$E^*_{\Lambda_S}=E^*_{\Lambda_L}$, but for the kaon, 
$E^*_{K_S}=E^*_{K_L}+m_K\delta m/E^*_{K_S}$. Also $\Gamma^*_{\Lambda}$ 
is the same for $\Lambda_L$ and $\Lambda_S$ in the $S^*$ frame of the
$\Lambda$, since 
the $\Lambda_L$ and $\Lambda_S$ move at the same speed in this frame.
In $S^*$ frame variables, then, the wave function, from (10), is
 
\begin{eqnarray}
\psi(x_{\Lambda},x_K)&=&\frac{1}{2}\left[{\rm exp}
\{i(p^*_{\Lambda_S}x^*_{\Lambda}
-E^*_{\Lambda_S}t^*_{\Lambda})-\frac{1}{2}\Gamma^*_{\Lambda}t^*_{\Lambda}+
i(p^*_{K_S}x^*_K-E^*_{K_S}t^*_K)-\frac{1}{2}\Gamma^*_St^*_K\}+\right.
\nonumber \\
&~~&~~~\left.{\rm exp}\{i(p^*_{\Lambda_L}x^*_{\Lambda}-
E^*_{\Lambda_L}t^*_{\Lambda})-
\frac{1}{2}\Gamma^*_{\Lambda}t^*_{\Lambda}+i(p^*_{K_L}x^*_K-E^*_{K_L}t^*_K)-
\frac{1}{2}\Gamma^*_Lt^*_K\}
\right]
\nonumber \\
&=&\frac{1}{2}\left[{\rm exp}
\{i(p^*_{\Lambda}x^*_{\Lambda}
-E^*_{\Lambda}t^*_{\Lambda})-\frac{1}{2}\Gamma^*_{\Lambda}t^*_{\Lambda}+
i(p^*_Kx^*_K-E^*_{K_S}t^*_K)-\frac{1}{2}\Gamma^*_St^*_K\}+\right.
\nonumber \\
&~~&~~~\left.{\rm exp}\{i(-p^*_{\Lambda}x^*_{\Lambda}-
E^*_{\Lambda}t^*_{\Lambda})-
\frac{1}{2}\Gamma^*_{\Lambda}t^*_{\Lambda}+i(-p^*_Kx^*_K-E^*_{K_L}t^*_K)-
\frac{1}{2}\Gamma^*_Lt^*_K\}
\right]
\end{eqnarray}
 
\noindent  where $(x^*_{\Lambda},t^*_{\Lambda})$ and $(x^*_K,t^*_K)$ 
are coordinates in the $S^*$ frame of each particle. The probability 
distribution is
 
\begin{eqnarray}
P(x^*_{\Lambda},x^*_K)&=&\frac{1}{4}{\rm exp}(-\Gamma^*_{\Lambda}
t^*_{\Lambda})\bigg[\left({\rm exp}(-\Gamma^*_{K_L}t^*_K)+
{\rm exp}(-\Gamma^*_{K_S}t^*_K)\right)+
\nonumber \\
&~&2~
{\rm exp}\left(-\frac{1}{2}(\Gamma^*_{KS}+\Gamma^*_{KL})t^*_K\right)
{\rm cos}\left(2p^*_{\Lambda}x^*_{\Lambda}+2p^*_Kx^*_K+
\frac{\delta (m^2)}{2E^*_K}t^*_K\right)\bigg]
\end{eqnarray}
 
\noindent For given $x^*_K$ and $t^*_K$, this distribution oscillates 
as a function of $x^*_{\Lambda}$, and therefore is an interference 
pattern that is stationary {\it in the $S^*$ frame of the} $\Lambda$. 
Now the frames $S_L$, $S_S$ and 
$\overline{S}$, defined by the velocities $\beta_L$, $\beta_S$ and 
$\overline{\beta}$, are almost the same as the $S^*$ frame, since 
$\delta (m^2)$ is small, but are not quite identical. Thus the frames 
$S_L$, $S_S$ and $\overline{S}$ move slowly in the frame $S^*$. As the 
$\Lambda$ moves out from the reaction point, each of these 
frames moves slowly across the interference pattern (16), giving the 
appearance of slow, long-wavelength oscillations. These oscillations 
are not real; they appear to be present if one looks only at the 
origin of a frame moving with the particle, and not at the full picture. 
In fact, eqn. (16) shows that there can be no 
oscillations of the $\Lambda$ or muon; the antinode in the $S^*$ frame,
defined by the reaction event, passes through all 
points on the trajectory of the $\Lambda$ or muon at some time. Thus 
there can be no stationary (in the c.m. frame) node at any point on 
the particle's path.
   
\vspace{0.1in}
 
Just as the above treatment confirms the absence of $\Lambda$ oscillations,
as derived in ref.[5], it also confirms the result of ref.[5] that the
wavelength of kaon oscillations is given by the usual formula. Two
expressions have been published that give different versions. Lipkin[10] 
showed that if the two mass eigenstates of the $K^0$ are regarded as 
having equal momentum, the oscillation wavelength may change by a factor of
2. However, the assumption of equal momentum is not correct for either of the
situations discussed here. Srivastava {\it et al.}[2] treat the kinematics
correctly but their error in the wavelength, which is greater than a
factor of 2, especially near threshold, results from incorrect
treatment of the transformations between the various rest frames
(see ref.[5]). The origins of these two factors have sometimes been 
confused in the 
literature (see [11,12,13] and sect.~V); we believe that our calculation 
(ref.[5] and the present paper) is the first to treat both the kinematics 
and the transformations between frames correctly.
 
\vspace{0.2in}
 
\centerline{\bf III. Wave packeting}
 
\vspace{0.2in}
 
As in any scattering process, it is implicitly assumed that the
plane-wave solutions discussed above will be used as a basis for the
construction of wave packets in order to correspond with a realistic
experimental situation (see, e.g., ref. [14]). In the case of 
oscillations, the use of wave packets is vital to the development of 
the interference pattern, and this section discusses the packeting
in some detail. We assume here that the size of the wave packets will 
be much larger than the separation of the centers of the $\Lambda_L$ 
and $\Lambda_S$ packets. If this is not the case, then the two packets 
will separate and coherence will be lost.

\vspace{0.1in}
 
We start with the 2-particle wave function for the final state
($\Lambda K$ or $\mu\nu$) from eqn. (10). The quantities $a_L$ and
$a_S$ are given by eqn. (4), but we write the phases in the exponents 
in the c.m. frame:
 
\begin{eqnarray}
\psi_{\Lambda K^0}(x_{\Lambda},x_{K^0})&=&
\nonumber \\
&~&\hspace{-1.0in}\frac{1}{2}\left[
{\rm exp}\left(i(p_{\Lambda_L}x_{\Lambda}-E_{\Lambda_L}t_{\Lambda})-
\frac{1}{2}\Gamma_{\Lambda}\tau_{\Lambda_L}+
i(p_{K_L}x_K-E_{K_L}t_K)-
\frac{1}{2}\Gamma_{K_L}\tau_{K_L}\right)\right.
\nonumber \\ 
&~~~&\hspace{-0.95in}+\left.
{\rm exp}\left(i(p_{\Lambda_S}x_{\Lambda}-E_{\Lambda_S}t_{\Lambda})-
\frac{1}{2}\Gamma_{\Lambda}\tau_{\Lambda_S}+
i(p_{K_S}x_K-E_{K_S}t_K)-
\frac{1}{2}\Gamma_{K_S}\tau_{K_S}\right)\right].
\end{eqnarray}
 
\noindent Following the usual procedure, we 
replace each sharp momentum, $p$, by $p+q$ where $q$ has a Gaussian 
distribution:
  
\begin{eqnarray}
\phi(q_{\Lambda})={\rm e}^{-q_{\Lambda}^2/2\sigma_{\Lambda}^2};
\hspace{0.5in}\phi(q_K)={\rm e}^{-q_K^2/2\sigma_K^2}.
\end{eqnarray}
 
\noindent The origin of the spread in the final-state momentum is
discussed in sect. IV. Normally, it will result from a measurement
on another particle in the system, possibly in the initial state, 
but that doesn't affect the argument.
  
\vspace{0.1in}
 
We apply (18) to each term in the wave function (17), giving
 
\begin{eqnarray}
\psi_{\Lambda K^0}(x_{\Lambda},x_{K^0})&=&
\nonumber \\
&~&\hspace{-0.6in}\frac{1}{2}\int\left\{{\rm exp}\left[i\left((p_{\Lambda_L}+
q_{\Lambda})x_{\Lambda}-(E_{\Lambda_L}+
q_{\Lambda}\frac{\partial E}{\partial p}\Bigg|_{p_{\Lambda_L}})
t_{\Lambda}\right)-\frac{1}{2}\Gamma_{\Lambda}\tau_{\Lambda_L}\right]
\times{\rm exp}\left[\Lambda\rightarrow K\right]\right.
\nonumber \\
&~&\hspace{-0.6in}\left.+{\rm exp}\left[i\left((p_{\Lambda_S}+
q_{\Lambda})x_{\Lambda}-(E_{\Lambda_S}+
q_{\Lambda}\frac{\partial E}{\partial p}\Bigg|_{p_{\Lambda_S}})
t_{\Lambda}\right)-\frac{1}{2}\Gamma_{\Lambda}\tau_{\Lambda_S}\right]
\times{\rm exp}\left[\Lambda\rightarrow K\right]\right\}
\nonumber \\
&~&\hspace{-0.6in}{\rm exp}\left(-q_{\Lambda}^2/2\sigma^2_{\Lambda}-
q_K^2/2\sigma^2_K\right)~~dq_{\Lambda}~~dq_K
\nonumber \\
&~&\hspace{-0.6in}=\frac{1}{2}
\left[{\rm exp}\left(i(p_{\Lambda_L}x_{\Lambda}-
E_{\Lambda_L}t_{\Lambda})-
\frac{1}{2}\Gamma_{\Lambda}\tau_{\Lambda_L}\right)
{\rm exp}\left(\frac{-\sigma^2}{2}(x_{\Lambda}-
\beta_{\Lambda_L}t_{\Lambda})^2\right)
{\rm exp}\left[\Lambda\rightarrow K\right]\right.
\nonumber \\
&~&\hspace{-0.6in}\left.+{\rm exp}\left(i(p_{\Lambda_S}x_{\Lambda}-
E_{\Lambda_S}t_{\Lambda})-
\frac{1}{2}\Gamma_{\Lambda}\tau_{\Lambda_S}\right)
{\rm exp}\left(\frac{-\sigma^2}{2}(x_{\Lambda}-
\beta_{\Lambda_S}t_{\Lambda})^2\right)
{\rm exp}\left[\Lambda\rightarrow K\right]\right],
\end{eqnarray}
 
\noindent where $\left[\Lambda\rightarrow K\right]$ denotes similar 
factors with $\Lambda$ replaced by $K$. The two Gaussian factors for 
$L$ and $S$ are indistinguishable in practical terms. Thus the 
$\Lambda$ and $K^0$ probability distribution that follows from this 
is the same as eqn. (11) for the plane-wave case except that it is 
multiplied by ${\rm exp}\left(-\sigma^{*2}_{\Lambda}x_{\Lambda}^{*2}-
\sigma^{*2}_Kx_K^{*2}\right)$. In the $S^*$ frame, the probability 
distribution (16) becomes
 
\begin{eqnarray}
P(x^*_{\Lambda},x^*_K)&=&\frac{1}{4}
{\rm exp}\left(-\sigma^{*2}_{\Lambda}x_{\Lambda}^{*2}-
\sigma^{*2}_Kx_K^{*2}\right)
{\rm exp}(-\Gamma^*_{\Lambda_L}t^*_{\Lambda})\bigg[
\left({\rm exp}(-\Gamma^*_{K_L}t^*_K)+{\rm exp}(-\Gamma^*_{K_S}t^*_K)\right)+
\nonumber \\
&~&2~
{\rm exp}\left(-\frac{1}{2}(\Gamma^*_{K_S}+\Gamma^*_{K_L})t^*_K\right)
{\rm cos}\left(2p^*_{\Lambda}x^*_{\Lambda}+2p^*_Kx^*_K+\frac{\delta(m^2)}
{2p_K}x_K\right)\bigg].
\end{eqnarray}
 
\noindent Here, $\sigma^* = \sigma/\gamma^*$ is the Gaussian momentum 
spread in the $S^*$ frames. Eqn. (20) then gives the probability, as a 
function of variables in the $S^*$ frames, for observation of a $\Lambda$ 
in conjunction with a neutral kaon with S = 1. As one might anticipate, 
the wave packet provides an 
envelope for the travelling oscillation patterns, but does not otherwise 
affect the analysis. As in eqn. (16), therefore, there are oscillatory
patterns in $x^*_{\Lambda}$ and $x^*_K$, and the patterns are
stationary in their respective $S^*$ frames. They can exist only if
the widths of the wave packets, $1/\sigma_{\Lambda}$ and $1/\sigma_K$,
are large enough that the pattern is not heavily damped away from the
points $x^*=0$. If the wave packets are narrow, so that 
$P(x^*_{\Lambda},x^*_K)$ is appreciable only when 
$p^*_{\Lambda},~p^*_K\sim 0$, then the particles will only be observed
essentially at their classical points. In this case, the argument of
the cosine term reduces to the value $(\delta(m^2)/2p_K)x_K$,
which is familiar from the standard treatment of strangeness
oscillations[1].
 
\vspace{0.2in}
 
\centerline{\bf IV. Observability of travelling oscillations}
 
\vspace{0.2in}
 
The discussion of section II shows that there is no stationary
oscillation pattern in the overall c.m. system for either of these
cases, to any order in $\delta (m^2)$. However, there is a pattern (eqn. (20))
that is stationary in the $S^*$ frame for the $\Lambda$ and we now examine 
the possibility of observing this pattern. Of course, the most important 
potential application of this would be the study of neutrino oscillations 
by measurements on the muon from the $\pi\rightarrow\mu\nu$ decay, but 
we also discuss the case of $\Lambda$s produced in the 
$\pi^-p\rightarrow\Lambda K^0$ reaction.

\vspace{0.1in}
 
The first important point is that this pattern, stationary in the 
$S^*$ frame of the recoiling muon or $\Lambda$, exists {\it only} if 
the neutrino flavour or the kaon strangeness is measured. This can 
be seen from eqn. (11) for the probability distribution of a 
$\Lambda$ recoiling against a $K^0$. If we had selected a $\bar{K^0}$ 
rather than a $K^0$, then the cosine term in eqn. (11) would have a 
minus sign. If we don't observe the strangeness of the kaon, then we 
must add these two probability distributions and the cosine term 
drops out.
 
\vspace{0.1in}
 
A further basic requirement is that it is necessary to measure 
coordinates for both particles in the final state. As can be seen 
in eqn. (20), both $x^*_{\Lambda}$ and $x^*_K$ occur in the argument of 
the cosine term; if either particle is unobserved, we must integrate 
eqn. (20) over it, and the oscillatory term in the $S^*$ frame vanishes. 
This is in contrast to the situation described by eqn. (11), where
particles are observed at their classical points, so that 
$x^*_{\Lambda}\sim x^*_K\sim 0$. Then, the argument of the cosine
is a function of $x^*_K$ but not of $x^*_{\Lambda}$, so that integration
over $x^*_{\Lambda}$ does not change the kaon oscillation pattern.
 
\vspace{0.1in}
 
To summarise, there are four requirements to observe the travelling 
oscillation pattern (20) in the $S^*$ frame:
 
\noindent (i) A detector to determine the kaon strangeness or the 
neutrino flavor. 
 
\noindent (ii) Detectors to measure both the time and position of 
each particle with appropriate resolution (see below). It is important 
to measure both time and position because the $S^*$ frame is moving, 
and it is necessary to know the position {\it in the} $S^*$ {\it frame} 
at which the particle is detected.
 
\noindent (iii) Some method to determine the position of the 
$S^*$ frame, since the observations must be transferred to that 
frame.
 
\noindent (iv) The $\Lambda$ oscillation pattern, eqn. (20) is centered 
on the production antinode which is stationary the $S^*$ frame, i.e. 
the classical particle position, 
and extends on either side of this by a distance determined by the 
spacial width of the wave packet. It is therefore necessary to prepare 
the state in a wave packet which is broad enough to cover a sufficient
width of the oscillation pattern.
 
\vspace{0.1in}
 
We can estimate the dimensions of the travelling interference pattern 
from eqn. (20). For the $\pi^-p\rightarrow\Lambda K^0$ reaction, we 
assume that the $\Lambda$s are produced in the at a center-of-mass
energy 0.2 GeV above threshold. Using the standard value of
$\delta m = 3.5\times 10^{-15}$GeV, the wavelength of $\Lambda$ 
oscillations in the $S^*$ frame is predicted to be 40 cm. For muons 
from the $\pi\rightarrow \mu\nu$ decay, we must make an assumption 
about the neutrino masses. If we assume $m_{\nu_{\mu}}\sim 10$eV and
$m_{\nu_e}\sim 0$eV, then the predicted neutrino oscillation wavelength
is about 50 cm. Smaller values for $m_{\nu_{\mu}}$ predict longer
oscillation wavelengths.
 
\vspace{0.1in}
 
The determination of the location of the $S^*$ frame is rather different
for the two cases. For the $\pi  p\rightarrow\Lambda K^0$ reaction, the 
necessary information could be provided by a detector in the pion beam, 
since the group velocities of all wave packets are known. This relies
on the fact that the reaction time is negligible, so that the time of
arrival of the pion at the proton is essentially the same as that when
the $S^*$ frame coincides with the overall c.m. frame. It would be 
necessary to measure positions to about 1 cm and times to about 1 ns or
better, for both the kaon and the $\Lambda$.
 
\vspace{0.1in}
 
Alternatively, a measurement of the kaon coordinates can also be used 
to locate the position of the antinode in the $S^*$ frame of the 
$\Lambda$, since this antinode is at 
the classical $\Lambda$ position which can be determined from the 
kaon coordinates. This measurement is required in any case for point 
(ii) above. In this case, measurement of the kaon narrows the wave
packets for both the kaon and the $\Lambda$ to widths determined
by the time resolution. Since the kaon wave packet is centered on the
point of observation, it follows that $x^*_K=0$. The position of the 
$S^*$ frame of the $\Lambda$ can readily be calculated from the kaon 
detection coordinates and the known classical velocities.
 
\vspace{0.1in}
  
The first of these methods will not work for the $\pi\rightarrow\mu\nu$ 
decay, since the pion lifetime is not negligibly small. However, the 
information could again be provided by the neutrino detector since we 
can be sure that the neutrino and  muon start from the pion decay point 
at the same time, and their velocities are known. 
 
\vspace{0.1in}
  
The other problem lies in the requirement (iv), the preparation of a 
quantum state with a sufficiently long wave packet. The problem is 
quite different for the $\Lambda K$ and $\mu\nu$ cases, since the 
state is prepared differently in these two cases. For the production 
of a $\Lambda K^0$ pair in the $\pi^-p\rightarrow\Lambda K^0$, the 
size of the final-state wave packets  is likely to be determined by 
the localisation in space-time of the incident pion. This may be 
determined by a detector in the beam. If so, the pion is localised 
in time by the time resolution, $t_{res}$, of the detector, and in 
space by $\beta_{\pi}ct_{res}$. Typically, a detector would have a 
time resolution of about $10^{-9}$ sec, which would give a spatial 
wave packet of about 30 cm. A wave packet this small would suppress 
the interference pattern of eqn. (20), since the Gaussian would have 
fallen off somewhat by the first zero in the oscillatory term. The
counter would have to be carefully designed to produce a coherent
wave packet over its time resolution, which would have to be
significantly longer than $10^{-9}$ sec. If there is no counter in
the incident beam, then the wave packeting will probably be 
produced by the properties of the accelerator. There seems to be
little discussion in the literature of the coherence length of
accelerator beams, and we know of no experimental measurements that
would determine whether a coherence length of several m is feasible.
It should be remarked that if the coherence length of the accelerator
beam is too short, then the use of a detector on the beam will 
not produce the desired wave packet, since the packet size will then
be driven by the accelerator coherence length rather than the
detector time resolution.
 
\vspace{0.1in}
 
In any case, the kaon detector is likely to be the limiting factor
that determines the packet size for the $\Lambda$. When the kaon is
detected, the system is prepared in a new quantum state in which the 
packet widths of both the kaon and the $\Lambda$ are determined by
the time resolution of the kaon detector.
 
\vspace{0.1in}
 
The case of muons from the $\pi\rightarrow \mu\nu$ decay is rather
different. The pion lifetime is $2.6\times 10^{-8}$ sec, and if there
are no measurements on the decay products and no limitations
on the observation time, this decay time will determine the packet
size. Again, however, the neutrino detector is likely to determine
the packet width for the muon, by preparing a state in which the 
muon position is defined to some accuracy, which will ultimately 
determine the coherence length of the muon.
 
\vspace{0.2in}
 
\centerline{\bf V. Discussion}
 
\vspace{0.2in}
 
We have shown that there are no stationary oscillations in the overall 
c.m. system of the recoil 
particle under any circumstances, in any order, and that the oscillations 
of the mixed particle, the $K^0$ or $\Lambda$, have a wavelength given by
the conventional expression. This is consistent with several other
recent treatments, but there has been some confusion in certain preprints
over two possible deviations from this expression for the wavelength.
Lipkin[10] pointed out that an error of a factor of 2 may result if the
two neutrino or kaon components are regarded (incorrectly) as having the
same momentum. A different error, which is always greater than 2, occurs 
in the work of Srivastava {\it et al.}[2,3,4]. As pointed out in section II, 
this is of quite a different origin, though Kiers and Weiss[11] and
Mohanty[12] seem to imply that they are the same. In fact, the
treatment of Kiers and Weiss[11] differs from ours in two ways. Firstly,
they take the source to be infinitely massive, so the kinematics are less
realistic. Also, their detector involves an inverse $\beta$ decay, and is
sharply resonant. This can give rise to detection effects which don't
occur with the non-resonant detectors assumed here.
 
\vspace{0.1in}
 
We have also shown that wave-packeting the states of the initial
particles produces the expected results on the final state. This is
in agreement with other publications, especially that of Grimus and 
Stockinger[15] who seem to be the first to discuss this in the recent
literature. Their treament
is more detailed and sophisticated than ours, but the result is
the same. Wave-packeting is also discussed by Giunti {\it et al.}[13],
whose results are again generally consistent with ours. In particular,
they point out, in agreement with our treatment and with Srivastava 
{\it et al.,} that the requirement of exact 4-momentum conservation at
the neutrino production point implies a mixture of 3-momenta for the
neutrino state.
 
\vspace{0.1in}
 
Although there is no {\it stationary} interference pattern for the recoil
(unmixed) particle, we have shown that, under the right circumstances, 
a travelling interference pattern should exist. This pattern is stationary
in a very specific frame, that in which the two momentum components of 
the recoil particle have equal energy. To observe this in the 
$\pi p \rightarrow\Lambda K^0$ case requires the appropriate time and 
position measurements to determine the location of this
frame and also a measurement of the strangeness of the neutral kaon. 
Further, it is necessary to prepare the initial state with a sufficiently
long coherence length. It may be possible to achieve all of these
requirements; probably a better understanding of the coherence length of
accelerator beams is needed to design an experiment. For the more
important case of a muon recoiling against a mixed neutrino, the 
experimental requirements are much more difficult to achieve. 
Unfortunately, it seems that the neutrino detector must be able to 
measure the neutrino flavour and also the time and position of the 
neutrino detection with the appropriate accuracy. Such a detector would 
presumably be capable of observing the neutrino oscillations directly; 
if so, the possiblity of inferring neutrino oscillations from measurements 
on the muon alone, which was one of the motivations for this work, would 
not be realised.
 
\vspace{0.1in}
 
%
%
 
\vspace{0.4in}
 
We are grateful to R.H. Dalitz, L. Okun and A. Zeilinger for useful 
discussions. We acknowledge support from the US DOE and the UK 
Rutherford Appleton Laboratory. 
 
\vspace{0.4in}
 
\noindent {\bf References}
 
\vspace{0.1in}
 
\noindent [1] See, e.g., the treatment in W.E. Burcham and M. Jobes,
Nuclear and Particle Physics (Longmans, Harlow, UK, 1995).

\vspace{0.1in}
 
\noindent [2] Y.N. Srivastava, A. Widom and E. Sassaroli, Phys. Lett. 
{\bf B344}, 436 (1995).
  
\vspace{0.1in}
 
\noindent [3] Y.N. Srivastava, A. Widom and E. Sassaroli, Zeitschrift
f\"{u}r Physik, {\bf C66}, 601 (1995).  
 
\vspace{0.1in}
 
\noindent [4] Y.N. Srivastava and A. Widom, preprints hep-ph/9707268,
hep-ph/9509261 and hep-ph/9612290.
  
\vspace{0.1in}
 
\noindent [5] J. Lowe, B. Bassalleck, H. Burkhardt, T. Goldman, A. Rusek
and G.J. Stephenson Jr., Phys. Lett. {\bf B 384}, 288 (1996).
 
\vspace{0.1in}
 
\noindent [6] T. Goldman, preprint hep-ph/9604357. 
 
\vspace{0.1in}
 
\noindent [7] F.\ Boehm and P.\ Vogel, {\it Physics of Massive
Neutrinos}, (Cambridge U.\ Press, 1987), p.\ 87.
 
\vspace{0.1in}
 
\noindent [8] A.D. Dolgov, A.Yu. Morozov, L.B. Okun and M.G. Schepkin,
Nucl. Phys. B{\bf 502}, 3 (1997).
  
\vspace{0.1in}
 
\noindent [9] B. Kayser, Proceedings of the Moriond Workshop on 
Electroweak Interactions and Unified Theories, Les Arcs, France, 
March 1995.
  
\vspace{0.1in}
 
\noindent [10] H. Lipkin, Phys. Lett. B {\bf 348}, 604 (1995); 
Y. Grossman and H. Lipkin, Phys. Rev. {\bf D 55}, 2760 (1997).
 
\vspace{0.1in}
 
\noindent [11] K. Kiers and N. Weiss, preprint hep-ph/9710289.
  
\vspace{0.1in}
 
\noindent [12] S. Mohanty, preprints hep-ph/9702248, hep-ph/9706328 
and hep-ph/9710284.
 
\vspace{0.1in}
 
\noindent [13] C. Giunti, C.W. Kim and U.W. Lee, preprint hep-ph/9709494;
C. Giunti and C.W. Kim, preprint hep-ph/9711363.
 
\vspace{0.1in}
 
\noindent [14] H. Burkhardt, Dispersion-relation Dynamics (North Holland,
Amsterdam, 1968), appendix A.

\vspace{0.1in}
 
\noindent [15]  W. Grimus and P. Stockinger, Phys. Rev. {\bf D54}, 3414
(1996).
 
%
 
\end{document}